\documentclass[aps,prb,twocolumn,showpacs,superscriptaddress]{revtex4}
\usepackage{bbm}
\usepackage{mathrsfs}
\usepackage{epsfig}
\usepackage{graphicx}
\usepackage{amsfonts}
\usepackage[figuresright]{rotating}
\usepackage{amssymb}
\usepackage{amsmath}
\usepackage{dcolumn}
\usepackage{bm}
\usepackage{psfrag}

\usepackage{natbib}

\newtheorem{theorem}{Theorem}
\newtheorem{lemma}{Lemma}

\newenvironment{proof}[1][Proof]{\noindent \textbf{#1: }}{\ \rule{0.5em}{0.5em}}

\newcommand{\COMMENTED}[1]{}

\def\avg#1{\langle#1\rangle}

\def\be{\begin{equation}} \def\ee{\end{equation}}
\def\bea{\begin{eqnarray}} \def\eea{\end{eqnarray}}

\def\nn{\nonumber}

\begin{document}
\title{Majorana Positivity and the Fermion Sign Problem
of Quantum Monte Carlo Simulations}
\author{Z. C. Wei}
\affiliation{Institute of Physics, Chinese Academy of Sciences,
P.O. Box 603, Beijing 100190, China}
\author{Congjun Wu}
\email{wucj@physics.ucsd.edu}
\affiliation{Department of Physics,  University of California,
San Diego, California 92093,  USA}
\author{Yi Li}
\affiliation{ Princeton Center for Theoretical  Science,  Princeton University,
Princeton,  NJ 08544}
\author{Shiwei Zhang}
\affiliation{Department of Physics, The College of William and Mary,
Williamsburg, Virginia 23187}
\author{T. Xiang}
\email{txiang@iphy.ac.cn}
\affiliation{Institute of Physics, Chinese Academy of Sciences, P.O. Box 603, 
Beijing 100190, China}
\affiliation{Collaborative Innovation Center of Quantum Matter, Beijing 100190, 
China}

\begin{abstract}
The sign problem is a major obstacle in quantum Monte Carlo simulations for 
many-body fermion systems. We examine this problem with a new perspective
based on the Majorana reflection positivity and Majorana Kramers positivity. 
Two sufficient conditions are proven for the absence of the fermion sign 
problem. 
Our proof provides a unified description for all 
the interacting lattice fermion models previously known to be free of the 
sign problem based on the auxiliary field quantum Monte Carlo method. 
It also allows us to identify a number of new sign-problem-free interacting 
fermion models including, but not limited to, lattice fermion models with 
repulsive interactions but without particle-hole symmetry and interacting
topological insulators with spin-flip terms.
\end{abstract}
\maketitle

\underline{Introduction:}
A major difficulty in the study of strongly correlated systems is the 
exponentially large many-body Hilbert spaces which are usually difficult 
to handle by analytic methods. 
Unbiased numerical methods are therefore indispensable. 
Among various numerical approaches, the quantum Monte Carlo (QMC) method 
can yield accurate results by taking stochastic but importance sampling 
over very small but representative portions of the many-body Hilbert 
space. 
An advantage of QMC is that it is scalable with the system size if there 
is no sign problem. 
Unfortunately, the sign problem exists in most interacting fermion and 
frustrated quantum spin systems.

The origin and manifestation of the sign problem vary in different QMC
algorithms. 
A frequently used algorithm for lattice fermions is the auxiliary
field determinantal method \cite{blankenbecler1981,hirsch1985}, in 
which the interaction terms are decoupled by the Hubbard-Stratonovich 
(HS) transformation into a superposition of quadratic fermion terms
in the background of imaginary-time-dependent auxiliary fields.
The fermion operators are then integrated out, yielding a fermion determinant 
which serves as the statistical weight for each HS field configuration. 
The sign problem emerges because this determinant is not always positive. 
In particular, the average value of the signs of these determinants often 
becomes exponentially small in the thermodynamic limit at low temperatures. 
This leads to uncontrollable statistical errors and ruins the calculation of QMC.
Although a great deal of efforts have been made to solve, at least partially, 
this problem \cite{zhangsw1999,zhangsw1999a, chandrasekharan1999, koonin1997, 
rombouts1998,imada2000,kashima2001,chandrasekharan2012,ogilvie2009,ogilvie2010},
a general solution is still lacking\cite{troyer2005}.

For certain classes of lattice fermion models, QMC simulations are proved
to be sign-problem-free. Familiar examples include the positive-$U$ Hubbard
model on a bipartite lattice at half-filling \cite{hirsch1985}, the
negative-$U$ Hubbard model \cite{hirsch1985, Scalettar1989}, and their
SU($2N$) generalizations \cite{caiZ2013,lang2013,wangD2014}.
The half-filled Kane-Mele-Hubbard model of interacting topological insulators is also sign-problem-free  \cite{zhengD2011,hohenadler2011}. 
For these models, after suitable HS decompositions with the Kramers time-reversal 
(TR) invariance \cite{hands2000,wu2005}, each determinant 
is factorized into a product of two complex
conjugate determinants defined in two subspaces with opposite spins.
A number of non-factorizable models, such as the multi-component and 
multi-band Hubbard models \cite{wu2003,capponi2004,berg2012}, 
the negative-$U$ Hubbard models with spin-orbit coupling \cite{tang2014,SHI2016},
can also be shown to be free of the sign problem. 
In these systems, instead of the determinant itself being factorizable,
the eigenvalues of the corresponding matrix are complex-conjugate paired, 
and real eigenvalues are doubly degenerate,
thus the determinant is non-negative valued. 
Recently, the Majorana HS decomposition was introduced in QMC simulations
\cite{lizx2015,lizx2015a}. 
It is applied to spinless fermion models with repulsive interactions at the 
particle-hole symmetric point. 
For some of the above mentioned models, the positivity of fermion determinants 
can be understood from the algebraic structure of the orthogonal split 
group $O(N,N)$ \cite{wangL2015,liuyh2015}.

The sign structures of the ground state wavefunctions of quantum lattice models 
are closely related to the reflection positivity of the Hamiltonian. 
The concept of reflection positivity was first introduced in the context 
of quantum field theory \cite{osterwalder1973}. 
Its main application in condensed matter physics started from Lieb's work 
on the spin reflection positivity in the Hubbard model 
\cite{Lieb1989,tianGS2004}, and was recently applied to systems 
with Majorana reflection positivity \cite{jaffe2015,jaffe2015b,weizc2015}.
For interacting fermion systems with these reflection positivities, it can 
be shown that their ground state wavefunctions are non-negative under 
certain suitably defined basis. 
However, these basis states are often non-local in real space, which is 
inconvenient for use in QMC simulations.

In this paper, we explore the QMC sign problem of lattice fermions from
the perspective of Majorana reflection positivity and Majorana
Kramers positivity defined below. We use the Majorana fermion
representation, because it allows any fermionic systems, whether
fermion number conserving or not, to be treated on equal footing.
In the framework of the determinantal QMC algorithm, the statistical
weight in the sampling is replaced by the trace of exponentials of
fermion bilinears resulting from the HS decomposition, which
is evaluated as a determinant.
A HS decomposition is said to be a ``positive decomposition'' if all
the generated determinants 
are positive semidefinite.
Below we show that there are at least two kinds of positive decompositions
which lead to QMC simulations free of the sign problem.
We dub them {\it Majorana reflection positive decomposition} and
{\it Majorana Kramers positive decomposition}, respectively.
These do not exhaust all positive decompositions.
They do, however, cover nearly all the interacting lattice fermion models
that are previously known to be sign-problem-free.
From these decompositions, we also identify a number of new
models which are free of the sign problem.

Let us begin with the determinantal QMC algorithm for a general lattice model 
of Dirac fermions. The Hamiltonian $H$ is a sum of a quadratic kinetic energy 
term $H_0$ and an interaction term of 4-fermion operators 
$H_I$ \cite{blankenbecler1981, hirsch1983, hirsch1985}. 
After the HS decomposition, the partition function $Z$ is expressed as
\begin{eqnarray}
Z & = & \mathrm{Tr} e^{-\beta H} = \lim_{M \to \infty}\sum_p \rho_p \\
\rho_p & = & \mathrm{Tr} \prod_{k=1}^{M}   e^{-\tau H_0}
e^{-\tau H_{I}(\eta_k)} ,
\label{eq:partition}
\end{eqnarray}
where $\beta$ is the inverse temperature, $\tau=\beta/M$ is the discrete
time interval, and $p=\{\eta_M(\{i\}), ..., \eta_k(\{i\}),  ...,
\eta_1(\{i\}) , i=1, \cdots, N\}$ represents a time-sequence of the
HS-field distributions, with $N$ the lattice size.
The decoupled interaction $H_I(\eta_k)$ contains only two-fermion terms,
and depends on the time-step size, $\tau$, and the spatial distribution
of the HS fields, $\eta_k(\{i\})$. The value of $\rho_p$ can be determined
by tracing out the fermion degrees of freedom in $H_0$ and $H_I$.
The formula for determining $\rho_p$ is given in Supplemental Material
(SM) 
\ref{app:rhop}.

At each lattice site, a Dirac fermion can be represented using two Majorana
fermions.
Thus the original $N$ Dirac fermions can be expressed in terms of $2N$
Majorana fermions. We divide these $2N$ Majorana fermions into two groups,
$\gamma^{(1)}_i$ and $\gamma^{(2)}_i$ ($1\le i \le N$), and define their
Clifford algebra operators as\cite{weizc2015}
\bea
\Gamma^{+}_\alpha&=&i^{[\frac{m}{2}]}\gamma^{(1)}_{i_1}...
\gamma^{(1)}_{i_m}, \ \ \,
\Gamma^{-}_\alpha =  (-i)^{[\frac{m}{2}]}\gamma^{(2)}_{i_1}...
\gamma^{(2)}_{i_m},
\eea
where $\alpha$ represents a sequence $\{i_1, i_2, ..., i_m\}$ with
$1\le i_1<...<i_m\le N$, and $[x]$ equals the largest integer less
than or equal to $x$. $\Gamma^{\pm}_\alpha$ is said to be even (odd)
if $m$ is even (odd).
The reflection operation $\theta$ is defined as an anti-linear automorphism
map:  $\theta(i)=-i$, $\theta(\gamma^{(1)}_i)=\gamma^{(2)}_i$ and 
$\theta(\gamma^{(2)}_i) =\gamma^{(1)}_i$.
Clearly, $\theta^2=1$ and $\theta(\Gamma^{\pm}_\alpha)=\Gamma^{\mp}_\alpha$.
A bosonic operator $O$ is Majorana reflection symmetric if $\theta(O)=O$,
and is Majorana reflection positive if it further satisfies the condition
\cite{jaffe2015,jaffe2015b}.
\bea
\mathrm{Tr}  [Q \circ \theta(Q) O]\ge 0,
\eea
where $Q=\sum_{\alpha} c_{\alpha}\Gamma_{\alpha}^+$ is an arbitrary operator
in the algebra spanned by the $\Gamma^+$-matrices
with $c_\alpha$'s the complex coefficients, and
$ Q\circ \theta(Q)=\sum_{\alpha\beta }c_{\alpha}c^*_{\beta}
\Gamma^+_{\alpha}\Gamma^-_{\beta}$.

In the Majorana representation, the bilinear terms in the expression
of $\rho_p$, including $H_0$ and $H_{I}(\tau_k)$, each can be expressed as
\begin{equation}
H_{bl} = \gamma^T V \gamma,
\label{eq:bl_majo}
\end{equation}
where $\gamma^T = (\gamma^{(1)}_i, \gamma^{(2)}_i)^T$ and $V$ is an
$2N\times 2N$ anti-symmetric matrix. $V$ is the coefficient matrix
of $H_0$ or $H_{I}(\tau_k)$ in the Majorana representation.

\underline{Majorana reflection positive decomposition: }
$V$ is defined as a Majorana reflection positive kernel
if it can be represented as
\begin{equation}
V=
\left(\begin{array}{cc}
A & iB\\
-iB^{T} & A^{*}
\end{array}\right),
\label{eq:bilinear}
\end{equation}
where $A$ and $B$ are $N\times N$ matrices. $A$ is complex anti-symmetric
satisfying $A^T=-A$. $B=B^\dagger$ is a Hermitian matrix which is either
positive semidefinite or negative semidefinite. ($B$ can be either positive
or negative semidefinite because, after a gauge transformation
$\gamma^{(2)}_j\to -\gamma^{(2)}_j$, $B$ becomes $-B$ and $A$
remains unchanged.)
A HS decomposition satisfying this condition will be called a
Majorana reflection positive decomposition.

{\theorem \label{theorem:reflection}
$\rho_p$ is positive semi-definite if all the coefficient matrices of
the bilinear fermion terms in Eq. (\ref{eq:partition}) are Majorana
reflection positive kernels}.

This theorem can be proved in two steps. 
The first is to show that, if $V$ is a Majorana reflection 
positive kernel, then $\exp (-\tau \gamma^T V \gamma)$ is
reflection positive. A proof on this was actually already given in
Ref. \cite{jaffe2015b}. This means that $\rho_p$ is just the trace
of a product of a series of reflection positive operators determined
by the exponentials of the bilinear fermion operators $H_0$ and
$H_I(\eta_k)$ in Eq. (\ref{eq:partition}).
The second step is to show that the product of a series of reflection
positive operators is also reflection positive and its trace is
non-negative.
A proof of this, as a lemma, is given in SM 
\ref{appendix:theorem1}.
Combining the above results, we have $\rho_p\ge 0$.
Thus the system is sign-problem-free in QMC simulations if all 
the kernels in Eq. (\ref{eq:partition}) are Majorana reflection positive.
The Majorana reflection can be regarded as
a generalization of the PT transformation
discussed in Refs \cite{bender1998, bender2010}.
However, this symmetry alone does not leads to
the positivity of $\rho_p$.
For example, 
if $B$ is not positive semidefinite, $\exp (-\tau \gamma^T V \gamma)$ remains
Majorana reflection symmetric, but is no longer Majorana reflection positive. 
In this case, $\rho_p$ is not always positive definite. 

Despite its seeming simplicity, Theorem \ref{theorem:reflection} covers
all two and higher dimensional interacting fermion models 
previously known to be sign-problem-free in determinantal QMC simulations, 
without imposing explicitly the TR-invariance in the HS decomposition.
These include the Hubbard model and its variations \cite{hirsch1985,
hohenadler2011,zhengD2011}, the interacting spinless fermion model
\cite{lizx2015,lizx2015a}, and other models whose coefficient matrices
in the Dirac fermion representation have the orthogonal split $O(N,N)$
group algebra structure \cite{wangL2015}.
Below we discuss two such examples, one
for spinless fermions and the other for spin one-half systems,
and prove they are sign problem free in new parameter regions
unknown before.

The first example is an interacting spinless fermion model defined on
a bipartite lattice. The model Hamiltonian is
\begin{eqnarray}
H_0&=&  -\sum_{i,j\in A} c^\dagger_i B_{1,ij} c_j
+\sum_{i,j\in B} c^\dagger_i B_{2,ij} c_j \nn \\
&+&\sum_{i\in A,j\in B} \left(c^\dagger_i F_{ij} c_j + h.c \right),
\label{eq:model1a}
\\
H_I &=& \sum_{ij} V_{ij} \left( n_i -\frac12 \right) \left( n_j -\frac12 \right),
\label{eq:model_1}
\end{eqnarray}
where $n_i = c^\dagger_i c_i$.
$B_{1}$ and $B_{2}$ are real symmetric matrices, both of which are
positive semidefinite (or, equivalently negative semidefinite).
$F$ is an arbitrary real matrix.
$V_{ij}\ge 0$ if $i$ and $j$ belong to different sublattices, and $V_{ij} \le 0$ otherwise.

$H_I$ can be decomposed into a bilinear form  by taking the following
HS transformation
\begin{equation}
  e^{-\tau V_{ij} \left(n_{i}-\frac{1}{2}\right) \left( n_{j}-\frac{1}{2} \right)} = \frac{1}{2} e^{-\frac{ \tau V_{ij} }{4}} \sum_{\eta=\pm} e^{\eta\lambda_{ij} (c^\dagger_{i} c_{j} + \nu c^\dagger_{j} c_{i})},
  \label{eq:HSspinless}
\end{equation}
where $\nu = -1$ if $i$ and $j$ belong to the same sublattices and  $\nu = +1$ otherwise.
In Eq.~(\ref{eq:HSspinless}), $\eta$ is a discrete local HS field, and
$\lambda_{ij}$ is determined by the equation $\sqrt{\nu}\,\lambda_{ij}
= \cosh^{-1} \exp (\tau V_{ij} /2)$.

It is simple to verify that both $H_0$ and the decoupled interaction terms in the exponent of Eq. (\ref{eq:HSspinless}) can be cast into the form,
\begin{eqnarray}
H_{bl}^\prime &=&  \sum_{i,j\in A} c^\dagger_i (C_{ij} -B_{1,ij}) c_j
+ \sum_{i,j\in B}  c^\dagger_i (D_{ij}+B_{2,ij}) c_j
\nn \\
&+& \sum_{i\in A,j\in B} \left(c^\dagger_i F_{ij} c_j + h.c \right),
\label{eq:corollary1}
\end{eqnarray}
where $C$ and $D$ are real anti-symmetric matrices satisfying $C=-C^T$ and
$D^T=-D$.
In SM-\ref{appendix:corollary1} A,
it is shown that the matrix kernel of
$H_{bl}^\prime$ is Majorana reflection positive. 
Therefore, the interacting spinless model $H_0+H_I$ is sign-problem-free,
according to Theorem \ref{theorem:reflection}.

If both of $B_1$ and $B_2$ vanish, the above interacting fermion
Hamiltonian defined on the honeycomb lattice is the model studied in Refs.
[\onlinecite{lizx2015,lizx2015a,wangL2015}].
It can be extended to include the on-site staggered chemical potential term,
and remains sign-problem free \cite{wangL2015,huffman2014}.
This is equivalent to only keeping the diagonal terms of $B_{1,2}$.
Generally speaking, in the presence of $B_{1,2}$, Eq. (\ref{eq:model1a}) and
Eq. (\ref{eq:model_1}) do not possess the particle-hole symmetry.
For example, consider the case
with $B_{1, ij}=\mu \delta_{ij}$ if $i\in A$ and $j\in A$,
and $B_{2,ij} =0$,
which is equivalent to applying
a uniform chemical potential $\mu /2$ and a staggered on-site potential
$(-)^i \mu / 2$ to the system. In the weak coupling limit, the single-particle spectrum splits into two bands and the band  gap is approximately
equal to $\mu/2$, which means that the chemical potential is located right at the bottom of the upper band. At zero temperature, the fermion density remains at half-filling. However, with increasing temperature, the fermion density begins to deviate from half-filling and the upper band is populated by fermions within an energy window of $T$ starting from the bottom of that band. This implies that a spinless fermion model with repulsive interactions can be simulated without the sign problem away from half filling.

Now let us consider a second example, a spin-1/2 fermion model with
Coulomb repulsion, spin-orbit coupling  and spin-flip terms,
again defined on a bipartite lattice.
This is a generalized Kane-Mele-Hubbard model.
The Hamiltonian $H=H_0+H_I$ is defined by
\begin{eqnarray}
H_0&=&-t\sum_{\avg{ij}\sigma} c^\dagger_{i\sigma} c_{j\sigma}
+i\lambda \sum_{\avg{\avg{ij}}\sigma }  \sigma c^\dagger_{i\sigma} c_{j\sigma}
\nonumber \\
&&+ \sum_{ij} (-)^j h_{ij} c^\dagger_{i\uparrow} c_{j\downarrow} + h.c.,
\label{eq:kanemele0} \\
H_I &=& U\sum_{i} (n_{i\uparrow}-\frac{1}{2})
(n_{i\downarrow}-\frac{1}{2}),
\label{eq:kanemele}
\end{eqnarray}
where $c_{i\sigma}$ ($\sigma=\uparrow$, $\downarrow$) is the annihilation operator of
fermion with spin $\sigma$. 
$h_{ij}$ is a real symmetric positive (or, negative) semi-definite matrix.
If we only keep the diagonal terms of $h_{ij}$, they reduce to
an in-plane staggered magnetic field distribution.
In the limit $\lambda = 0$ and $h_{ij}=0$, this Hamiltonian becomes the half-filled
Hubbard model.
For finite $\lambda$ and $h_{ij}$, it breaks the SU(2) invariance.
In the absence of $h_{ij}$, the $z$-component of total spin $S_z$
remains conserved.
In this case, Eq. (\ref{eq:kanemele}) is known to be sign-problem-free
\cite{zhengD2011,hohenadler2011}.
However, in the presence of $h_{ij}$, both the $S_z$ conservation and
the TR symmetry are broken.
The previous proof for the absence of the sign problem is no longer
valid \cite{zhengD2011,hohenadler2011}.

To show the above model is sign-problem free,
let us first consider the following bilinear Hamiltonian of spin-1/2 fermions,
\begin{eqnarray}
H_{bl}^{\prime\prime} &=& \sum_{ij} \left(c^\dagger_{i\uparrow} M_{ij} c_{j\uparrow}
+c^\dagger_{i\downarrow} M^*_{ij} c_{j\downarrow} \right)
\nonumber \\
&& - \sum_{ij} h_{ij}
\left( c^\dagger_{i\uparrow} c^\dagger_{j\downarrow} +c_{j\downarrow} c_{i\uparrow}\right),
\label{eq:corollary2}
\end{eqnarray}
where $M_{ij}$ is an arbitrary $N\times N$ complex matrix.
It can be shown that $H_{bl}^{\prime\prime}$ is Majorana reflection positive.
A proof of this is given in SM-\ref{appendix:corollary1} B.

The Coulomb interaction $H_I$ can be decomposed into a bilinear form
by the following HS transformation,
\begin{equation}
e^{-\tau U (n_{i\uparrow}-\frac{1}{2})(n_{i\downarrow}-\frac{1}{2})}=
 \frac{1}{2} e^{\frac{1}{4} \tau U} \sum_{\eta = \pm}
e^{i\lambda^\prime\eta (n_{i\uparrow}+n_{i\downarrow}-1)},
\label{eq:Hubbard}
\end{equation}
where $\lambda^\prime=\cos^{-1} \exp (-\tau U/2)$.
By taking a particle-hole transformation for the down-spin fermion operators
$c^\dagger_{j\downarrow} \rightarrow (-)^j c_{j\downarrow}$ and keeping the up-spin
fermion operators unchanged, the bilinear exponent on the right hand
side of Eq. (\ref{eq:Hubbard}) becomes
$i\lambda^\prime\eta (n_{i\uparrow} - n_{i\downarrow})$.
Under the same transformation, the $t$- and $\lambda$-terms in $H_0$ defined by
Eq. (\ref{eq:kanemele0}) remain unchanged, but the $h$-term becomes the
second term of Eq. (\ref{eq:corollary2}).
Thus $H_0$ is also Majorana reflection positive, and the Kane-Mele-Hubbard model
defined in Eq. (\ref{eq:kanemele}) is free of the QMC sign
problem according to Theorem \ref{theorem:reflection}.

The absence of the sign problem of the Hamiltonian defined by
Eqs (\ref{eq:kanemele0}) and (\ref{eq:kanemele}) is actually beyond the
framework of TR-invariant decompositions in the Dirac fermion
representation \cite{wu2005,hands2000}. It provides an opportunity to
study the effect of TR-symmetry breaking in two-dimensional interacting
topological insulators through QMC simulations \cite{kane2005,wu2006}.
The $h$-terms, which flip electron spins, can arise from the scattering
of magnetic impurities.
The magnetic impurities on the edges of a two-dimensional topological
insulator can destabilize the helical edge states by opening gaps.
The interplay among interaction effects, band structure topology,
and magnetic impurities, is an interesting topic that deserves further
investigation.

\underline{Majorana Kramers positive decomposition: } 
We next present a second theorem for the absence of sign problem
based on the Kramers symmetry structure of Majorana fermions.
{\theorem  \label{theorem:symp}
$\rho_p$ is non-negative if there exist two transformation operators,
$S$ and $P$, such that
\begin{eqnarray}
    \label{eq:TR_MJ} S^T V S &=&  V^{*}, \\
    \label{eq:TR_MJ_2} P V P^{-1}&=&V,
\end{eqnarray}
where $S$ is a real antisymmetric matrix satisfying $S^2=-I$ and $S^T=-S$,
$P$ is a symmetric or antisymmetric Hermitian matrix satisfying $P^2=I$,
and $P$ anti-commutes with $S$,  i.e., $PS =-SP$.
}

A proof of this theorem is given in SM-\ref{app:theorem2}. 
The HS decomposition satisfying Eqs. (\ref{eq:TR_MJ})
and (\ref{eq:TR_MJ_2})
is termed as Majorana Kramers positive decomposition.
It is a generalization of the Kramers TR-invariant decomposition used
in the determinant QMC of Dirac fermions in Refs.
[\onlinecite{wu2005,hands2000}].
Here $S$, combined with complex conjugation $C$, defines an anti-unitary
Kramers transformation operator $T=SC$ satisfying $T^2=-1$.
A kernel $V$ that satisfies Eqs. (\ref{eq:TR_MJ}) and
(\ref{eq:TR_MJ_2}) is also invariant under the anti-linear
transformation $T^\prime=PSC$.
If $P$ is antisymmetric, then $T^\prime$ is also a Kramers operator, 
satisfying $(T^\prime)^2=-1$, and $V$ is not Majorana reflection symmetric.
On the other hand, if $P$ is symmetric, then $(T^\prime)^2=1$.
In this case, $V$ is both Majorana reflection and Kramers symmetric.

Eq. (\ref{eq:TR_MJ}) ensures that the Majorana coefficient matrix $V$
is symmetric under the TR transformation.
But the symmetry alone does not assure $\rho_p$ to be non-negative.
This is because the eigenvalues of $\prod_{k}\exp (-\tau V_k)$ always appear in
pairs: If $\Lambda_\alpha$ is an eigenvalue, so is $\Lambda_\alpha^{-1}$.
Moreover,  from Eq. (\ref{eq:TR_MJ}), it can be shown that $\Lambda^*_\alpha$
and $(\Lambda_\alpha^*)^{-1}$ are also eigenvalues.
If $\Lambda_\alpha$ is modulus 1, then $\Lambda_\alpha^*=\Lambda_\alpha^{-1}$.
In this case, these four eigenvalues reduce to two if $\Lambda_\alpha$
is not doubly degenerate.
According to the expression of $\rho_p$ in terms of
$\Lambda_\alpha$'s in SM-\ref{app:rhop},
$\rho_p$ may not be positive-definite.
The condition defined by Eq. (\ref{eq:TR_MJ_2}) adds an extra
constraint to the Majorana coefficient matrix $V$. 
It enforces the double degeneracy of the eigenvalues of  
$\prod_{k}\exp (-\tau V_k)$ 
when they are modulus one.
This Kramers degeneracy assures $\rho_p\ge 0$.

Theorem \ref{theorem:symp} is valid independent of the specific
representations of $S$ and $P$. This implies that there is significant
flexibility in choosing the HS decomposition scheme.
Below we consider some simple realizations of $S$ and $P$ operators.
Assuming the system contains $N=2L$ sites, we label the lattice sites by
two indices $(a,i)$ with $a=1,2$ ($a$ can be also regarded as the index
for the orbital degrees of freedom) and $i=1,..., L$, and the
corresponding Majorana fermion operators by $\gamma^{(\mu)}_{a,i}$ with
$\mu=1,2$ the index of the two Majorana fermions at each site.
Operator $S$ can then be taken as $S=i\sigma_2\otimes \tau_0\otimes I$,
where $I$ is the identity matrix in the sector of $i$, and
$\sigma_\alpha$ and $\tau_\alpha$ denote the identity ($\alpha =0$) and
Pauli ($\alpha = 1, 2, 3$) matrices in the sectors of $\mu$ and $a$,
respectively.
The role of $S$ is to map $\gamma^{(1)}_{a,i}$ to $\gamma^{(2)}_{a,i}$
and $\gamma^{(2)}_{a,i}$ to $-\gamma^{(1)}_{a,i}$.

$P$ can be either anti-symmetric or symmetric.
An  anti-symmetric  $P$ can be defined as $\sigma_{1,3}\otimes
\tau_2\otimes I$. $S$ and $P$ thus defined can be applied to the interacting
fermion model investigated in Ref. \cite{hands2000,wu2005}.
For a symmetric $P$, there are more choices, including any of the following,
$\sigma_{1}\otimes \tau_\alpha \otimes I$ and $\sigma_{3}\otimes
\tau_\alpha \otimes I$ ($\alpha = 0, 1, 3$).
For each of these, there exists a corresponding class of $V$'s satisfying
Eqs. (\ref{eq:TR_MJ}) and (\ref{eq:TR_MJ_2}).
For example for $P=\sigma_1\otimes \tau_0\otimes I$, $V$ can be generally
expressed as
\begin{equation}
V =  \sum_{\alpha} \left( \sigma_0\otimes \tau_\alpha \otimes A_\alpha
+ i\sigma_1\otimes \tau_\alpha \otimes B_\alpha \right),
\label{eq:tr-v}
\end{equation}
where $A_\alpha$ and $B_\alpha$ with $\alpha =0,1,$ and $3$ are real
anti-symmetric matrices, and $A_2$ and $B_2$ are imaginary symmetric matrices.
The interacting fermion models that can be decomposed into the form of
Eq. (\ref{eq:tr-v}) would represent a new class of models without
the QMC sign-problem.
An example of sign-problem free Hamiltonian that satisfies
Eq. (\ref{eq:tr-v}) is given in SM \ref{SM:exp_Th2}.

\underline{Summary:}
We have shown that interacting fermion models are free of the sign
problem in determinantal QMC simulations if the bilinear Hamiltonians
obtained with the HS-decomposition possess the Majorana reflection
positivity or the Majorana Kramers positivity.
The two theorems we have proven cover all the sign-problem-free
interacting lattice models that are previously known.
It also allows us to identify a number of new interacting fermion
models without the QMC sign problem.

{\it Acknowledgments}
We thank L. Wang for helpful discussions.
Z. C. W. and T. X. are supported by the National Natural Science Foundation of China (Grants No. 11190024 and No. 11474331) and by the National Basic Research Program of China (Grants No. 2011CB309703). C. W. is supported by the NSF DMR-1410375 and AFOSR FA9550-14-1-0168.
C. W. acknowledges the CAS/SAFEA International Partnership Program
for Creative Research Teams of China, and the President's Research
Catalyst Awards CA-15-327861 from the University of California
Office of the President.
Y.L. thanks the Princeton Center for Theoretical Science at Princeton
University for support.
S.Z. is supported by the NSF (Grant no.~DMR-1409510) and the Simons Foundation.


\appendix

\section{Determination of $\rho_p$}
\label{app:rhop}

$\rho_p$ defined in Eq. (2) can be determined from the determinant of the
coefficient matrices of $H_0$ and $H_I(\tau_k)$.
In the Dirac fermion representation, $H_0$ and $H_I(\eta_k)$ can be
generally expressed as
\begin{eqnarray}
    H_0 & = & \sum_{i,j} c^\dagger_i h_{0,ij} c_j, \\
    H_{I}(\eta_k) &= & \sum_{ij} c^\dagger_i h_{I,ij}(\eta_k)c_j ,
\end{eqnarray}
where $h_0$ and $h_I(\eta_k)$ are the coefficient matrices for the corresponding bilinear interactions.
It can be shown that,
after tracing over fermions, $\rho_p$ is given by the following determinant \cite{hirsch1985}
\begin{equation}
  \rho_p = \det\left(I+ \prod_{k=1}^M e^{-\tau h^0} e^{-\eta h^I (\eta_k)} \right).
\end{equation}

The above expression holds if the fermion number is conserved. When
the fermion number is not conserved, it is more convenient to evaluate $\rho_p$ in the Majorana representation. If we use $V_0$ and $V_I(\eta_k)$ to represent the Majorana coefficient matrices of $H_0$ and $H_I(\eta_k)$, then $\rho_p$ is given by
\begin{equation}
    \rho_p = \left[ \det  \left(I + A\right) \right]^{1/2} ,
\end{equation}
where
\begin{equation}
    A=\prod^M_{k=1} e^{-4 \tau V^0} e^{-4 \eta V^I(\eta_k)}. \label{eq:ortho}
\end{equation}
$V^0$ and $V^I(\eta_k)$ are anti-symmetric matrices. $A$ belongs to the special orthogonal group SO$(2N,C)$.
Its eigenvalues are pairwised as $(\Lambda_i,\Lambda_i^{-1})$, and
thus the fermion Pfaffian $\rho_p$ can be expressed as
\begin{equation}
\rho_p 
=\prod_{i=1}^{N} \left(1 + \Lambda_i \right).
\label{eq:pf}
\end{equation}

\section{Lemma}
\label{appendix:theorem1}

In this part, we prove the following Lemma which is used in the
proof of Theorem 1 in the main text.
{\lemma  The product of a series of reflection positive operators,
$X=\prod_{k=1}^M O_k$, is reflection positive.
Its trace is non-negative, i.e., $\mathrm{Tr}  X\ge 0$.
}

\proof
We express each operator $O_k$ in the form of Eq. (4), and
diagonalize its coefficient matrices as $W^{\mu}(k)=U^\mu(k) \Lambda^\mu(k) U^{\mu,\dagger}(k)$ ($\mu=o,\, e$), where $\Lambda^\mu(k)$ is a positive semidefinite diagonal matrix.
We then have
\[
O_k=  \sum_{\gamma}A_{\gamma}^{e+}(k) [A_{\gamma}^{e-}(k)]^*
+ i \sum_{\gamma} A_\gamma^{o+}(k) [A_\gamma^{o-}(k)]^*,
\]
where
\begin{eqnarray}
A^{\mu +}_{\gamma}(k)&=&\sum_{\alpha}
\Gamma^{+}_\alpha [U^\mu(k)]_{\alpha\gamma} [\Lambda^\mu_\gamma(k)]^{\frac{1}{2}},
\nonumber \\
(A^{\mu -}_\gamma(k))^*&=&\sum_{\alpha}
\Gamma^{-}_\alpha [U^\mu(k)]_{\alpha\gamma}^* [\Lambda^\mu_\gamma(k)]^{\frac{1}{2}}.
\nn
\end{eqnarray}

$X$ can now be expressed as
\begin{eqnarray}
X&=&\sum_{r} i^{m_{r}} \prod_{k}
\left\{ A^{\mu_k +}_{\gamma_k} (k) [A^{\mu_k -}_{\gamma_k}(k)]^* \right\},
\label{eq:X}
\end{eqnarray}
where $r$ represents a configuration $\{\gamma_k,\mu_k\}$, and $m_r$ is the total number of $A^{o +}$'s.

Grouping $m_r$ by parity, we reorganize Eq. (\ref{eq:X}) as
\begin{eqnarray}
X= \sum_{m_r\in even} B^{e,+}_{r}[B^{e,-}_{r}]^*
+i \sum_{m_r\in odd} B^{o,+}_{r}[B^{o,-}_{r}]^*,
\end{eqnarray}
where $B^{\nu,\pm}_{r}=\prod_{k=1}^M A^{\mu_k \pm}_{\gamma_k} (k)$
for $\nu = e$ and $o$.

Let us expand $B_r^{\nu,\pm}$ as
\begin{eqnarray}
B_r^{e(o),+}&=& \sum_{\alpha\in even (odd)} b_{r,\alpha} \Gamma^+_\alpha,
\nn \\
(B_r^{e(o),-})^*&=& \sum_{\alpha\in even (odd)} b^*_{r,\alpha} \Gamma^-_\alpha.
\end{eqnarray}
Consider a general operator $Q\in {\cal U^+}$ with
$Q=\sum_{\alpha} c_\alpha \Gamma^+_\alpha$, we have
\begin{equation}
\mathrm{Tr} [Q\circ \theta(Q) X] =\sum_r \sum_{\alpha}
|b_{r,\alpha} c_\alpha|^2\ge 0,
\end{equation}
where the parity of $\alpha$ in the second summation
follows that of $m_r$, and $\circ$ is defined as
\begin{equation}
Q\circ \theta(Q)=\sum_{\alpha\beta \in \text{even}}c_{\alpha}c^*_{\beta}
\Gamma^+_{\alpha}\Gamma^-_{\beta}
+i \sum_{\alpha\beta \in \text{odd}} c_{\alpha}c^*_{\beta}
\Gamma^+_{\alpha}\Gamma^-_{\beta}.
\end{equation}
From Ref. [\onlinecite{jaffe2015b}], we know that $X$ is reflection positive.

By setting $Q=1$, we have $\mathrm{Tr}  X\ge 0$.
More explicitly, we expand
$B_r^{e,+}= D^{-1} w_{r} +\sum^\prime_\alpha c_\alpha \Gamma_\alpha^{+}$ with
$w_r=\mathrm{Tr} B_r^{e,+}$, where the identity matrix is excluded in the sum. Therefore
\begin{eqnarray}
\mathrm{Tr}  X= D^{-1} \sum_{r\in even} |w_{r}|^2\ge 0.
\end{eqnarray}
\textit{Q.E.D.}

\section{Majorana reflection positivity}
\label{appendix:corollary1}

The Majorana reflection operator $\theta$ is anti-linear, satisfying the 
equation  $\theta^2=1$, and can be viewed as a generalized $PT$ operator 
of Majorana fermions \cite{bender1998, bender2010}. 
It is simple to show that $\rho_p$ is real if all the matrix kernels defined in Eq. (2) in the main text are Majorana reflection symmetric. However, as these kernels defined at different imaginary times generally do not commute with each other, the reflection symmetry alone does not guarantee $\rho_P$ to be positive definite. To ensure the positivity of $\rho_P$, these kernels, as stated in Theorem 1, must be positive defined. The minus-sign problems exists, for example, in the Hamiltonian defined by Eq. (7) in the main text if $B_1 = -B_2 = \mu\not=0$, even though all the matrix kernels remain Majorana reflection symmetric.

Below we show that the bilinear interactions defined by Eqs. (10) and (13) in
the main text are Majorana reflection positive.

\subsection{Interaction defined by Eq. (10)}
\label{app:spinless}

Let us take the following Majorana representation for the fermion operators
\begin{eqnarray}
c_i & = & \frac{1}{2}\left(\gamma_{1i}-i\gamma_{2i} \right), \qquad i \in A ,
\\
c_i & = & \frac{1}{2}\left(\gamma_{3i}-i\gamma_{4i} \right), \qquad i \in B .
\end{eqnarray}
Defining 
\begin{equation}
  \gamma^{(1)}_i=\left(
    \begin{array}{c}    \gamma_{1i} \\  \gamma_{4i} \end{array}\right),
  \qquad
  \gamma^{(2)}_i=\left(
    \begin{array}{c} \gamma_{2i}\\ \gamma_{3i} \end{array}\right),
\end{equation}
we can then convert the bilinear interaction $H^\prime_{bl}$ given in Eq. (10) into the form of Eq. (6) with the coefficient matrix defined by
\begin{eqnarray}
A&=&\frac{1}{2}\left(\begin{array}{cc}
 C & -iF\\
iF^{T} & D
\end{array}\right),
\\
B&=&\frac{1}{2}\left(\begin{array}{cc}
B_1 &0 \\
0   &B_2 \\
\end{array}
\right),
\end{eqnarray}
where $B_{1,2}$ are positive (negative) semidefinite real symmetric matrices.
Furthermore, it can be shown that $A$ is complex antisymmetric and $B$ is Hermitian. Thus $H_{bl}^\prime$ is Majorana reflection positive.

\subsection{Interaction defined by Eq. (13)}
\label{app:spinful}

For the interaction defined by Eq. (13), we use the following Majorana
representation of fermion operators
\begin{eqnarray}
c_{i\uparrow} & = & \frac{1}{2}\left(\gamma_{1i}-i\gamma_{2i}\right),
\\
c_{i\downarrow}& = & \frac{1}{2}\left(\gamma_{3i} - i\gamma_{4i}\right).
\end{eqnarray}
Eq. (13) can then be transformed into the standard form of Eq. (6) by defining
\begin{equation}
    \gamma^{(1)}_i = \left(
        \begin{array}{c} \gamma_{1i} \\  \gamma_{2i} \end{array}\right),
        \qquad
    \gamma^{(2)}_i =\left(
        \begin{array}{c} \gamma_{4i} \\ \gamma_{3i} \end{array}\right).
\end{equation}
The coefficient matrix of $H_{bl}^{\prime\prime}$ in the Majorana representation is defined by
\begin{eqnarray}
    A & = & \frac{1}{2}\left(\begin{array}{cc}
            M_a & -iM_s\\
            iM_s &  M_a
            \end{array}\right),
    \\
    B & = & \frac{1}{2}\left(\begin{array}{cc}
            h& 0 \\
            0&  
h \\
\end{array}\right),
\end{eqnarray}
where $M_{s} = (M+M^T)/2 $ and $M_a= (M-M^T)/2$ are the symmetric and
antisymmetric parts of $M$, respectively; $h$ is real symmetric
and positive (or negative) semidefinite.
As $B$ is positive or negative semidefinite, the interaction defined by Eq. (13)
is Majorana reflection positive.

\section{Proof of Theorem 2}
\label{app:theorem2}

$S$ combined with complex conjugation defines an anti-unitary TR transformation $T$ satisfying $T^2=-1$.
As $P^2 = 1$, the eigenvalues of $P$ take two values, $\pm 1$. We denote by $\chi_\alpha$ ($\alpha = 1, ... N$) the columnar eigenvectors of $P$ with eigenvalue $+1$, i.e.,
\begin{equation}
    P\chi_\alpha = \chi_\alpha.
\end{equation}
As stated in Theorem 2, $P$ can be either anti-symmetric or symmetric. Below we consider these two cases separately.

1. $P$ is anti-symmetric

In this case, $P$ is purely imaginary. For every eigenvector $\chi_\alpha$ of eigenvalue $+1$, it is simple to show that $\chi_\alpha^*$ is an eigenvector of $P$ with eigenvalue $-1$,
\begin{equation}
    P \chi_\alpha^* = - \chi_\alpha^*.
\end{equation}
Using these eigenvectors, we define the Dirac fermion operators as
\begin{eqnarray}
    c^\dagger_\alpha & = & \frac{1}{\sqrt{2}} \gamma^T \chi_\alpha , \\
    c_\alpha & = & \frac{1}{\sqrt{2}} \gamma^T \chi_\alpha^* .
\end{eqnarray}
As $P$ commutes with the Majorana coefficient matrix $V$, the fermion number operator, defined by
\begin{equation}
    C=\frac{1}{4}\gamma^T P \gamma =\sum_\alpha  (c^\dagger_\alpha c_\alpha -\frac{1}{2}),
\end{equation}
is conserved by the bilinear Hamiltonian defined in Eq. (6). Furthermore, it is straightforward to show that $S\chi_\alpha^*$ is an eigenvector of $P$ with eigenvalue $+1$,
\begin{equation}
    P\left( S\chi_\alpha^* \right) = -S\left( P\chi_\alpha^* \right)= S\chi_\alpha^*.
\end{equation}
Thus the fermion operator $c_\alpha$ after the TR transformation
\begin{equation}
   T c^\dagger_\alpha T^{-1}= \sum_{n} (S\chi^{(m,*)})_n \gamma_n \equiv c^\dagger_{\bar\alpha}
\end{equation}
remains a fermion operator denoted as $c^\dagger_{\bar \alpha}$. This means that the fermion number is also conserved by the TR transformation. Thus $H_{bl}$ defined by Eq. (5) is a TR-invariant decomposition
in terms of Dirac fermions, and $\rho_p$ is non-negative,
which is proven in Ref. \cite{hands2000,wu2005}.

2. $P$ is symmetric

In this case, $P$ is purely real and cannot be used to build up a conserved quantity. Its eigenvectors can be also set real. For each given eigenvector $\chi_\alpha$ of $P$ with eigenvalue $+1$, we have
\begin{equation}
    P (S\chi_\alpha)=-S\chi_\alpha\,.
\end{equation}
Thus $S\chi_\alpha$ is an eigenvector with eigenvalue $-1$. $O=\left( \chi_1,...\chi_N , S\chi_1,... S\chi_N \right)$ forms a $2N\times 2N$ real orthogonal transformation matrix $O$. It transforms the coefficient matrix $V$ into a block diagonal form
\begin{equation}
    O^T V O= \left( \begin{array}{cc}
        X& 0 \\ 0& X^* \end{array} \right),
\label{eq:tran_1}
\end{equation}
where $X$ is an $N\times N$ anti-symmetric matrix whose matrix elements
are defined by $X_{\alpha \alpha^\prime} = \chi_\alpha^T  V \chi_{\alpha^\prime}$.
Thus after a global real orthogonal transformation $O$, all the coefficient
matrices can be cast into the form of Eq. (6) with zero off-diagonal blocks,
which is Majorana reflection positive. From Theorem 1, we know that
$\rho_p$ is positive semidefinite.

\section{An example of Theorem 2}
\label{SM:exp_Th2}
As an application of Theorem 2, let us consider a spinless interaction
lattice fermion model $H=H_0+H_I$,
\bea
H_0&=& \sum_{ij} \Delta_{ij} \Big\{ c_i c_j +c^\dagger_j c^\dagger_i\Big\},
\nn \\
H_I&=& \sum_{ij} V_{ij}(n_i-\frac{1}{2})(n_j-\frac{1}{2}),
\eea
where $\Delta_{ij}$ are real anti-symmetric matrix, and $V_{ij}<0$.
Here the fermion number is only conserved module 2.
After taking the Hubbard-Stratonovich decomposition using Eq. (9)
in the main text, the resultant fermion bilinear
forms can be expressed as follows
\bea
H_{bl}= \sum_{ij} c^\dagger_i C_{ij} c_j +\sum_{ij}
\Delta_{ij} \Big\{ c_i c_j +c^\dagger_j c^\dagger_i\Big\}.
\label{eq:bl_theorem2}
\eea
By defining $c_i=\frac{1}{2}(\gamma_{1,i} +i \gamma_{2,i})$,
Eq. (\ref{eq:bl_theorem2}) can be cast into the form of
$H_{bl}=\gamma^T V \gamma$ with $\gamma^T=(\gamma^{(1)}_i,\gamma^{(2)}_i)$
and
\bea
V= \left( \begin{array}{cc}
A_0 & iB_0 \\
iB_0& A_0 \\
\end{array}
\right),
\eea
where $A_{0,ij}=C_{ij}$, and $B_{0,ij}=\Delta_{ij}$.
This matrix kernel $V$ is a special case of Eq. (17) in the main text
which contains only terms with $\alpha=0$.

\section{Relationship between Theorem 1 and Theorem 2}
\label{appendix:PT}
Theorem 1 and Theorem 2 cover different classes of models. 
But they have certain overlaps. The overlap happens if $P$ defined in 
Theorem 2 is symmetric, or, 
equivalently the off-diagonal block matrix $B$ in the 
matrix kernel $V$ defined in Eq. (6) for Theorem 1
in the main text, vanishes. 
In this case, the matrix kernel $V$ is both Majorana reflection 
symmetric and Majorana Kramers symmetric, and satisfies the conditions 
for both Theorems simultaneously. If the $P$ operator is anti-symmetric, 
i.e.  imaginary, it can be shown the matrix kernel $V$ that satisfies 
the condition of Theorem 2 is not Majorana reflection symmetric, because 
there is not an anti-linear transformation $\theta$ satisfying $\theta^2=1$ 
under which $V$ is invariant. It can be also shown that a reflection 
symmetric kernel $V$ that satisfies the condition of Theorem 1 is 
not Kramers symmetric if its off-diagonal block matrix $B$ is finite.


\begin{thebibliography}{10}

\bibitem{blankenbecler1981}
R. Blankenbecler, D.~J. Scalapino, and R.~L. Sugar, Phys. Rev. D {\bf 24},
  2278  (1981).

\bibitem{hirsch1985}
J.~E. Hirsch, Phys. Rev. B {\bf 31},  4403  (1985).

\bibitem{zhangsw1999}
S.~W. {Zhang}, eprint arXiv:cond-mat/9909090, in `Quantum Monte Carlo Methods
  in Physics and Chemistry,' Ed. M.P. Nightingale and C.J. Umrigar (Kluwer
  Academic, Dordrecht, 1999  (1999).

\bibitem{zhangsw1999a}
S.~W. Zhang, Phys. Rev. Lett. {\bf 83},  2777  (1999).

\bibitem{chandrasekharan1999}
S. Chandrasekharan and U.~J. Wiese, Phys. Rev. Lett. {\bf 83},  3116  (1999).

\bibitem{koonin1997}
S.~E. Koonin, D.~J. Dean, and K. Langanke, Phys. Rep. {\bf 278},  1  (1997).

\bibitem{rombouts1998}
S. Rombouts, K. Heyde, and N. Jachowicz, Phys. Rev. C {\bf 58},  3295  (1998).

\bibitem{imada2000}
M. Imada and T. Kashima, J. Phys. Soc. Jpn. {\bf 69},  2723  (2000).

\bibitem{kashima2001}
T. Kashima and M. Imada, J. Phys. Soc. Jpn. {\bf 70},  2287  (2001).

\bibitem{chandrasekharan2012}
S. Chandrasekharan and A. Li, Phys. Rev. D {\bf 85},  091502  (2012).

\bibitem{ogilvie2009}
M.~C. {Ogilvie} and P.~N. {Meisinger}, SIGMA {\bf 5},  047  (2009).

\bibitem{ogilvie2010}
M.~C. Ogilvie, P.~N. Meisinger, and T.~D. Wiser, Int. J. Theo. Phys. {\bf 50},
  1042  (2010).

\bibitem{troyer2005}
M. {Troyer} and U.-J. {Wiese}, Phys. Rev. Lett. {\bf 94},  170201  (2005).

\bibitem{Scalettar1989}
R.~T. Scalettar {\it et~al.}, Phys. Rev. Lett. {\bf 62},  1407  (1989).

\bibitem{caiZ2013}
Z. Cai {\it et~al.}, Phys. Rev. Lett. {\bf 110},  220401  (2013).

\bibitem{lang2013}
T.~C. Lang {\it et~al.}, Phys. Rev. Lett. {\bf 111},  066401  (2013).

\bibitem{wangD2014}
D. Wang {\it et~al.}, Phys. Rev. Lett. {\bf 112},  156403  (2014).

\bibitem{zhengD2011}
D. Zheng, G.-M. Zhang, and C. Wu, Phys. Rev. B {\bf 84},  205121  (2011).

\bibitem{hohenadler2011}
M. Hohenadler, T.~C. Lang, and F.~F. Assaad, Phys. Rev. Lett. {\bf 106},
  100403  (2011).

\bibitem{hands2000}
S. Hands {\it et~al.}, Eur. Phys. J. C {\bf 17},  285  (2000).

\bibitem{wu2005}
C. Wu and S.-C. Zhang, Phys. Rev. B {\bf 71},  155115  (2005).

\bibitem{wu2003}
C. Wu, J.~P. Hu, and S.~C. Zhang, Phys. Rev. Lett. {\bf 91},  186402  (2003).

\bibitem{capponi2004}
S. Capponi, C. Wu, and S.~C. Zhang, Phys. Rev. B {\bf 70},  220505(R)  (2004).

\bibitem{berg2012}
E. {Berg}, M.~A. {Metlitski}, and S. {Sachdev}, Science {\bf 338},  1606
  (2012).

\bibitem{tang2014}
H.-K. {Tang}, X. {Yang}, J. {Sun}, and H.-Q. {Lin}, EPL (Europhysics Letters)
  {\bf 107},  40003  (2014).

\bibitem{SHI2016}
H. Shi, P. Rosenberg, S. Chiesa, and S.~W. Zhang, to be published  (2016).

\bibitem{lizx2015}
Z.-X. Li, Y.-F. Jiang, and H. Yao, Phys. Rev. B {\bf 91},  241117  (2015).

\bibitem{lizx2015a}
Z.-X. {Li}, Y.-F. {Jiang}, and H. {Yao}, New Journal of Physics {\bf 17},
  085003  (2015).

\bibitem{wangL2015}
L. {Wang} {\it et~al.}, Phys. Rev. Lett. {\bf 115},  250601  (2015).

\bibitem{liuyh2015}
Y.-H. Liu and L. Wang, Phys. Rev. B {\bf 92},  235129  (2015).

\bibitem{osterwalder1973}
K. Osterwalder and R. Schrader, Comm. Math. Phys. {\bf 31},  83  (1973).

\bibitem{Lieb1989}
E.~H. Lieb, Phys. Rev. Lett. {\bf 62},  1201  (1989).

\bibitem{tianGS2004}
G. Tian, Journal of Statistical Physics {\bf 116},  629  (2004).

\bibitem{jaffe2015}
A. {Jaffe} and F.~L. {Pedrocchi}, Annales Henri Poincar{\'e} {\bf 16},  189
  (2015).

\bibitem{jaffe2015b}
A. {Jaffe} and B. {Janssens}, Arxiv:1506.04197  .

\bibitem{weizc2015}
Z.-C. Wei, X.-J. Han, Z.-Y. Xie, and T. Xiang, Phys. Rev. B {\bf 92},  161105
  (2015).

\bibitem{hirsch1983}
J.~E. Hirsch, Phys. Rev. B {\bf 28},  4059  (1983).

\bibitem{bender1998}
C.~M. Bender and S. Boettcher, Phys. Rev. Lett. {\bf 80},  5243  (1998).

\bibitem{bender2010}
C.~M. {Bender} and P.~D. {Mannheim}, Physics Letters A {\bf 374},  1616
  (2010).

\bibitem{huffman2014}
E.~F. Huffman and S. Chandrasekharan, Phys. Rev. B {\bf 89},  111101  (2014).

\bibitem{kane2005}
C.~L. Kane and E.~J. Mele, Physical Review Letters {\bf 95},  146802  (2005).

\bibitem{wu2006}
C. Wu, B.~A. Bernevig, and S.-C. Zhang, Phys. Rev. Lett. {\bf 96},  106401
  (2006).

\end{thebibliography}

\end{document}